\documentclass[showpacs,amsmath,amssymb, prb,twocolumn]{revtex4}
\usepackage{graphicx}
\usepackage{dcolumn}
\usepackage{bm}

\begin{document}

\title{Amplification phenomena of Casimir force fluctuations on close scatterers coupled via a coherent fermionic fluid}

\author{F. Romeo$^{1,2}$} 
\affiliation{$^{1}$Dipartimento di Fisica "E.R. Caianiello", Universit\`a di Salerno, I-84084 Fisciano (SA), Italy\\
$^{2}$CNR-SPIN Salerno, I-84084 Fisciano (SA), Italy}

\date{\today}
\begin{abstract}
We study the mechanical actions affecting close scatterers immersed in a coherent fermionic fluid. Using a scattering field theory, we theoretically analyse the single-scatterer and the two-scatterer case. Concerning the single-scatterer case, we find that a net force affects the scatterer dynamics only in non-equilibrium condition, i.e. imposing the presence of a non-vanishing particle current flowing through the system. The force fluctuation (variance) is instead not negligible both in equilibrium and in non-equilibrium conditions. Concerning the two-scatterer case, an attractive fluid-mediated Casimir force is experienced by the scatterers at small spatial separation, while a decaying attractive/repulsive behavior as a function of the scatterer separation is found. Furthermore, the Casimir force fluctuations acting on a given scatterer in close vicinity of the other present an oscillating behavior reaching a long distance limit comparable to the value of the single-scatterer case. The relevance of these findings is discussed in connection with fluctuation phenomena in low-dimensional nanostructures and cold atoms systems.
\end{abstract}


\maketitle
\section{Introduction}
When two conducting neutral plates are placed at micrometric distance in a vacuum, a net force between them arises. This force, first theoretically described by Casimir\cite{casimir,mostepanenko-book} and experimentally confirmed by several experiments \cite{casimir-exp}, originates from the boundary conditions imposed to the virtual photons of the electromagnetic field and it is the direct manifestation of the existence of the zero-point energy of a quantum field. Further studies have demonstrated that Casimir force can be either attractive or repulsive depending on the geometry of conducting plates \cite{casimir-geometry}. These studies have received great attention, since Casimir interaction can affect the correct operativity of nanoelectromechanical systems causing (or preventing) frictions and adhesion phenomena\cite{capasso2009}.\\
The original problem of the mediated interaction of two plates immersed in a virtual photons gas (bosonic fluid) can be generalized to the case of a fermionic quantum fluid. Along this line, the problem of the mediated interaction between two scattering centers (impurities) immersed in a one dimensional fermionic fluid has been studied both in interacting\cite{zwerger2007,recatiPRA05} and non-interacting case\cite{zhabinskaya2008}. The interest in these studies comes from the fact that in particularly clean systems, like one-dimensional ultracold atomic gases, individually trapped atoms, playing the role of impurities, can be created and controlled\cite{recati2005}. Artificial defects, eventually movable, can be also created encapsulating neutral fullerenes in single-wall carbon nanotubes\cite{ferone2010}. Alternatively, nanotubes functionalization using chemical adsorbates can be performed to create localized defects whose features can be probed within a field effect transistor setup \cite{nanotube-ads}. In all these systems, Casimir interaction is reminiscent of the peculiarities of the quantum fluid which can be probed looking at the impurity dynamics. However, impurities immersed in a quantum fluid are subject to Casimir force and its fluctuations, the latter being produced by the quantum nature of the fluid-mediated interaction. Fluctuations of Casimir force have received limited attention in literature\cite{casimir-fluct2010} and further efforts are needed in this direction.\\
To fill this vacancy, in this work, we formulate a scattering field theory of the mechanical actions experienced by scattering centers immersed in a coherent fermionic fluid. Non-equilibrium, non-perturbative scattering effects, decoherence, interaction with boundaries, Casimir-like interaction and its fluctuations can be studied within the proposed theoretical framework. We provide a comprehensive treatment of fluid-mediated forces coming from a quantum fluid confined in one dimension, while many-body interaction effects are not included in the theory. Due to the structure of the theory, divergence problems sometimes affecting alternative formulations are not present in our approach. As a relevant results of the theory, we show that force fluctuations on close scatterers immersed in a coherent fluid are correlated by Casimir interaction. We discuss this finding in connection with fluctuation phenomena in solids.\\
The organization of the paper is the following. In Sec.~\ref{sec:mech-act} we present the mechanical actions on a scatterer immersed in a quantum liquid with special emphasis on the momentum transfer equation. The force operator and the force fluctuation operator are also introduced. In Sec.~\ref{sec:scattering-obs} observables are derived within the scattering field theory. In particular, in Sec.~\ref{sec:force-single}, we derive the force acting on a single impurity under equilibrium and out-of-equilibrium conditions. We also derive the equilibrium force fluctuations acting on the scatterer. In Sec.~\ref{sec:force-two} we present the two-scatterer case deriving the Casimir-like force and its fluctuation. The role of decoherence is also discussed. Conclusions are given in Sec.~\ref{sec:concl}. The S- and T-matrix of a Dirac delta potential, the thermal averages of the scattering fields, the T-matrix model of random dephasing and the M-matrix derivation are reported in the appendices \ref{app: A}, \ref{app: B}, \ref{app: C}, \ref{app: D}, respectively.

\section{Mechanical actions on a scatterer immersed in a coherent fermionic fluid}
\label{sec:mech-act}
We consider a coherent fermionic fluid confined to one dimension and described by the second quantization field $\hat{\Psi}(x)$ whose evolution is determined by the Hamiltonian:
\begin{eqnarray}
\hat{H}_0=\int dx \hat{\Psi}^{\dagger}(x)\Bigl[-\frac{\hbar^2 \partial^{2}_x}{2m}\Bigl]\hat{\Psi}(x),
\end{eqnarray}
where we disregard the spin degree of freedom which is unessential for our purposes. When an impurity is introduced in the system, the particle density is perturbed by the scatterer potential $V(x)$ and thus the additional term $\hat{H}_I=\int dx \hat{\Psi}^{\dagger}(x)V(x)\hat{\Psi}(x)$ is added to the unperturbed fluid Hamiltonian $\hat{H}_0$, being the total Hamiltonian $\hat{H}=\hat{H}_0+\hat{H}_I$. The Heisenberg equation of motion of the field $\hat{\Psi}(x)$, i.e. $i\hbar \partial_t \hat{\Psi}(x)=[\hat{\Psi}(x),\hat{H}]$, complemented by the fermionic anticommutation relations $\{\hat{\Psi}(x),\hat{\Psi}^{\dagger}(x')\}=\delta(x-x')$ and $\{\hat{\Psi}(x),\hat{\Psi}(x')\}=\{\hat{\Psi}^{\dagger}(x),\hat{\Psi}^{\dagger}(x')\}=0$, completely determines the space/time evolution of the fermionic field according to the Schr\"{o}dinger-like equation:
\begin{eqnarray}
i\hbar \partial_t \hat{\Psi}(x)=-\frac{\hbar^2 \partial^{2}_x}{2m} \hat{\Psi}(x)+V(x)\hat{\Psi}(x).
\end{eqnarray}
In order to characterize the mechanical interactions between the fluid and the scatterer we need to study the momentum transfer equation. To this end, let us introduce the momentum density operator $\hat{\rho}_P$ which is defined in terms of fermionic field as follows:
\begin{eqnarray}
\hat{\rho}_P=-\frac{i \hbar}{2}\Big[\hat{\Psi}^{\dagger}(x)\partial_x \hat{\Psi}(x)-\partial_x \hat{\Psi}^{\dagger}(x)\hat{\Psi}(x) \Big].
\end{eqnarray}
From a classical viewpoint, the linear momentum of the center of mass of a mechanical system is not conserved in the presence of a net force. According to this observation, in general, the momentum density balance of the quantum fluid cannot be expressed in the form of a continuity equation and sink/source terms are expected. This expectation is easily verified by direct computation of the time derivative of $\hat{\rho}_P$ which leads to the following momentum density balance equation \cite{yurke90,presilla92}:
\begin{eqnarray}
\label{eq:momentum-balance}
\partial_{t}\hat{\rho}_P+\partial_x \hat{J}_P=-\hat{\Psi}^{\dagger}(x)\partial_x V(x) \hat{\Psi}(x),
\end{eqnarray}
where we have introduced the momentum current density operator:
\begin{eqnarray}
\hat{J}_P&=&\frac{\hbar^2}{4m}\Big \{2\partial_x \hat{\Psi}^{\dagger}(x)\partial_x\hat{\Psi}(x)-\hat{\Psi}^{\dagger}(x)\partial^{2}_x\hat{\Psi}(x) \nonumber \\
&-&\partial^{2}_x\hat{\Psi}^{\dagger}(x)\hat{\Psi}(x) \Big \}.
\end{eqnarray}
The r.h.s. of Equation~(\ref{eq:momentum-balance}), i.e. $\hat{f}(x)=-\hat{\Psi}^{\dagger}(x)\partial_x V(x) \hat{\Psi}(x)$, represents the force density operator (sink/source term) responsible for a momentum transfer from the potential $V(x)$ to the coherent fluid. An opposite mechanical action is exerted by the fluid on the scatterer. The force operator $\hat{\mathcal{F}}$ related to the fluid pressure acting on the scatterer (originating the potential $V(x)$) can be computed as:
\begin{eqnarray}
\label{eq:force-op-barr}
\hat{\mathcal{F}}&=&\int_{\Omega} dx (-\hat{f}(x)) \nonumber \\
&=&\int_{\Omega} dx \hat{\Psi}^{\dagger}(x)\partial_x V(x) \hat{\Psi}(x),
\end{eqnarray}
where the integration domain is restricted to a subset $\Omega$ of the real axis centered at the scatterer position $\bar{x}$ and fulfilling the requirement $\partial_x V(x)\neq 0$.\\
In the following we specialize our treatment to describe rectangular potentials which represent a convenient approximation of a local impurity potential. For the rectangular potential $V(x)=U \theta(x-x_1)\theta(x_2-x)$ ($x_2>x_1$ and $\theta(x)$ the Heaviside step function) centered at $\bar{x}=(x_1+x_2)/2$ we obtain
\begin{equation}
\label{eq:grad-rect}
\partial_x V(x)=U \delta(x-x_1)-U\delta(x-x_2),
\end{equation}
with $\delta(x)$ the Dirac delta function.
Thus, using Eq.~(\ref{eq:force-op-barr}) and Eq.~(\ref{eq:grad-rect}) with the integration domain $\Omega \equiv [x_1,x_2]$, the force operator $\hat{\mathcal{F}}$ can be presented in the following simple form:
\begin{equation}
\hat{\mathcal{F}}=U\Big[\hat{\Psi}^{\dagger}(x_1)\hat{\Psi}(x_1)-\hat{\Psi}^{\dagger}(x_2)\hat{\Psi}(x_2) \Big],
\end{equation}
while its expectation value $\langle \hat{\mathcal{F}}\rangle$ has to be evaluated performing a quantum statistical average accounting for the occupation of the relevant degrees of freedom.\\
The quantum nature of the fermionic fluid allows force fluctuations which can be evaluated introducing the force fluctuation operator $\delta \hat{\mathcal{F}}(t)=\hat{\mathcal{F}}(t)-\langle \hat{\mathcal{F}}\rangle$, whose quantum average is a vanishing quantity by construction. The time correlation of the force fluctuation $\delta \hat{\mathcal{F}}(t)$ can be studied introducing the noise correlator:
\begin{eqnarray}
K(t,t')=\frac{1}{2}\langle\{\delta \hat{\mathcal{F}}(t),\delta \hat{\mathcal{F}}(t')\}\rangle,
\end{eqnarray}
which is the symmetrized version of the quantity $\langle \delta \hat{\mathcal{F}}(t)\delta \hat{\mathcal{F}}(t') \rangle=\langle\hat{\mathcal{F}}(t)\hat{\mathcal{F}}(t')\rangle-\langle\hat{\mathcal{F}}\rangle^2$. Finally the noise spectral density $K(\Omega)$ can be obtained as
\begin{equation}
K(\Omega)=\int d \tau K(t+\tau,t) \exp(i \Omega \tau).
\end{equation}

\section{Force and force fluctuations within the scattering field theory}
\label{sec:scattering-obs}
We have introduced the force operator $\hat{\mathcal{F}}$ and the force fluctuation operator $\delta \hat{\mathcal{F}}(t)$ in second quantization. However, the characterization of  the system properties requires the computation of expectation values such as $\langle \hat{\mathcal{F}}\rangle$ or $\langle \delta \hat{\mathcal{F}}(t)\delta \hat{\mathcal{F}}(t') \rangle$. Hereafter, we develop a scattering field theory \`{a} la B\"{u}ttiker\cite{buttiker92} which allows the computation of the net force acting on a scatterer and the force fluctuations under equilibrium and non-equilibrium conditions. Two cases are analyzed: (i) the single-scatterer case in which the fermionic fluid is perturbed by only one impurity; (ii) the two-scatterer case in which the fluid is affected by the simultaneous perturbation originated by two scattering centers.

\subsection{Single-scatterer case}
\label{sec:force-single}
\begin{figure}[!h]
\includegraphics[clip,scale=0.6]{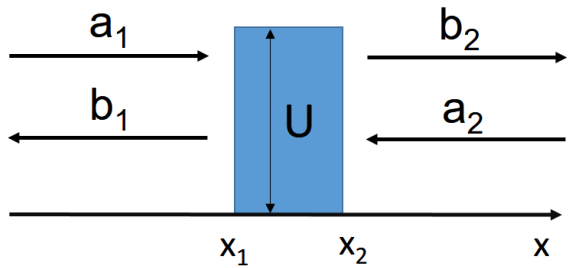}
\caption{(color online) Rectangular potential $V(x)$ centered at $\bar{x}=(x_1+x_2)/2$. Incoming ($a_i$) and outgoing ($b_i$) scattering operators are used to decompose the fermionic field $\hat{\Psi}(x)$ on the left ($x<x_1$, $i=1$) and on the right ($x>x_2$, $i=2$) side of the scatterer.}
\label{fig:fig1}
\end{figure}
Let us consider the mechanical actions exerted by the fermionic fluid on a single scatterer (see Fig.~\ref{fig:fig1}). The fermionic field $\hat{\Psi}(x,t)$ can be decomposed using the scattering field operators $\hat{a}_i(E)$ and $\hat{b}_i(E)$. In particular the annihilation operator $\hat{a}_i(E)$ destroys an incoming particle having energy $E$ generated in the \textit{i}th lead and moving towards the scattering center, while $\hat{b}_i(E)$ is the annihilation operator of an outgoing processes going from the scattering center towards the \textit{i}th lead. In the scattering language the lead 1 represents the region on the left side of the scatterer ($x<x_1$), while the lead 2 is located on the right side of the scattering center ($x>x_2$). Thus, the scattering fields $\hat{\Psi}_{i}(x,t)$ describing the fluid on the left or right side of the scatterer are written as
\begin{eqnarray}
\label{eq:scattering-fields}
\hat{\Psi}_{1}(x,t)=\int \frac{dE e^{-i\frac{Et}{\hbar}}}{\sqrt{h v(E)}}[\hat{a}_1(E) e^{i k_E x}+\hat{b}_1(E) e^{-i k_E x}]\nonumber \\
\hat{\Psi}_{2}(x,t)=\int \frac{dE e^{-i\frac{Et}{\hbar}}}{\sqrt{h v(E)}}[\hat{a}_2(E) e^{-i k_E x}+\hat{b}_2(E) e^{i k_E x}],
\end{eqnarray}
where we have introduced the group velocity $v(E)=\hbar k_E/m$ and the wavevector $k_E=\sqrt{2mE}/\hbar$ of the scattering modes. The incoming and the outgoing annihilation operators are related by the unitary scattering matrix $S_{ij}(E)$ according to the relation $\hat{b}_i(E)=\sum_j S_{ij}(E)\hat{a}_j(E)$, while the lead correlation functions are completely determined by the quantum statistical average $\langle \hat{a}^{\dagger}_{i}(E)\hat{a}_{j}(E')\rangle=\delta_{ij}\delta(E-E')f_{i}(E)$. The Fermi distribution $f_{i}(E)$ defines the thermodynamic occupation of a quantum state of energy $E$ in the \textit{i}th lead by means of a local electrochemical potential $\mu_i$ and a local effective temperature $T_i$. Within the scattering field formalism the expectation value $\langle \hat{\mathcal{F}} \rangle$ is computed as
\begin{equation}
\label{eq:scattering-force-single}
\langle \hat{\mathcal{F}} \rangle=U\Big[\langle\hat{\Psi}_{1}^{\dagger}(x_1)\hat{\Psi}_{1}(x_1)\rangle-\langle\hat{\Psi}_{2}^{\dagger}(x_2)\hat{\Psi}_{2}(x_2)\rangle \Big].
\end{equation}
Equation~(\ref{eq:scattering-force-single}) can be explicitly evaluated in terms of the reflection probability of the (rectangular) scatterer potential $|r(E)|^2=|S_{11}(E)|^2=|S_{22}(E)|^2$ according to the following expression:
\begin{equation}
\label{eq:force-res-singl}
\langle \hat{\mathcal{F}} \rangle=\frac{2U}{h}\int \frac{dE}{v(E)}|r(E)|^2(f_{1}(E)-f_{2}(E)).
\end{equation}
where $f_i(E)$ is the Fermi distribution of the \textit{i}th lead. Equation~(\ref{eq:force-res-singl}) shows that under equilibrium condition ($f_1(E)=f_2(E)$) or for reflectionless scattering potentials ($|r(E)|=0$) no net force is applied on the scatterer. On the other hand, assuming a weak voltage bias defined by the electrochemical potentials $\mu_i=\mu-(-)^{i+1}|e|V_{b}/2$ ($i \in \{1, 2\}$) and low temperature, i.e. $T_1=T_2=T \approx 0$, one obtains the following linear response result\cite{note1}:
\begin{equation}
\label{eq:linear-resp-single-barrier}
\langle \hat{\mathcal{F}} \rangle=-\frac{|e|U}{\pi \hbar v_F}|r(E=\mu)|^2 V_{b},
\end{equation}
where $-|e|$ represents the electron charge, $v_F$ is the Fermi velocity, while $\mu \approx m v_F^{2}/2$ is the equilibrium chemical potential of the system. The non-equilibrium force given in Eq.~(\ref{eq:linear-resp-single-barrier}) represents the pressure exerted by the electron wind generated by the voltage bias $V_{b}$ and provides a positive force on the scatterer when the applied voltage is negative. An expression similar to Eq.~(\ref{eq:linear-resp-single-barrier}) can be derived under the assumption of a thermal gradient applied to the fluid ($T_1 \neq T_2$) showing that a thermal imbalance originates mechanical stress on the scattering center.\\
Since we are interested in describing ultralocal potentials $V(x)$ originated, for instance, by reticular defects in a solid, we now specialize our result (Eq.~(\ref{eq:linear-resp-single-barrier})) to the case of very small barrier width $x_2-x_1$. Under this condition the original rectangular potential can be approximated by the Dirac delta potential $V(x)=2 a U\delta(x-\bar{x})$, with $2a=x_2-x_1$ and $\bar{x}=(x_1+x_2)/2$. Here the local approximation of the original potential is designed to meet the condition $\int_{x_1}^{x_2}dx V(x)=2aU$, which is also respected by the original potential. Under these assumptions, the force expectation value $\langle \hat{\mathcal{F}} \rangle$ given in Eq.~(\ref{eq:scattering-force-single}) has to be computed taking the limit
$x_1, x_2 \rightarrow \bar{x}$, being the limit well defined within the scattering approach. Thus the force acting on an ultralocal potential can be written in the simple form:
\begin{equation}
\langle \hat{\mathcal{F}} \rangle \approx -\frac{|e| V_{b} }{\pi \hbar v_F}\frac{U^3}{U^2+(\frac{\hbar v_F}{2a})^2},
\end{equation}
where we used the scattering matrix derived in appendix \ref{app: A}.
In solid state systems the non-equilibrium force due to the electron wind on the barrier can be compensated by elastic deformations of the lattice determining a new equilibrium position of the impurity site. On the other hand, if $\langle \hat{\mathcal{F}}\rangle$ is strong enough to overcome the lattice forces, the impurity electromigration phenomenon occurs consisting in a permanent damaging of the lattice.\\
We now analyse the force fluctuation acting on the impurity potential in the absence of a net force, i.e. assuming an equilibrium condition ($V_{b}=0$). Since we are interested in the steady-state behavior we focus our attention on the low-frequency part of the force fluctuation spectrum. To this end, we have to compute the quantity:
\begin{eqnarray}
K_{0}=\lim_{\Omega \rightarrow 0} \int d(t-t')\exp(i\Omega(t-t'))\langle \delta \hat{\mathcal{F}}(t) \delta \hat{\mathcal{F}}(t')\rangle.
\end{eqnarray}
Hereafter we outline the analytical derivation of the quantity $K_{0}$ within the scattering field theory. Starting from Equation~(\ref{eq:scattering-force-single}) (under the assumption of ultralocal potential) and using the scattering fields, the force operator $\hat{\mathcal{F}}(t)$ can be written in the form:
\begin{equation}
\label{eq:scattering-force-op}
\hat{\mathcal{F}}(t)=\int dE dE' e^{i \frac{(E-E')}{\hbar}t}\sum_{ij}\Gamma_{ij}(E,E')\hat{a}^{\dagger}_{i}(E)\hat{a}_{j}(E'),
\end{equation}
where the integration kernel $\Gamma_{ij}(E,E')=U[\Lambda^{(1)}_{ij}(E,E')-\Lambda^{(2)}_{ij}(E,E')]$ is related to the scattering matrix of the local potential through the auxiliary functions:
\begin{equation}
\Lambda^{(l)}_{ij}(E,E')=\frac{\delta_{il}\delta_{jl}+S^{\ast}_{li}(E)S_{lj}(E')}{h \sqrt{v(E)v(E')}}.
\end{equation}
Furthermore, the force expectation value $\langle \hat{\mathcal{F}}\rangle$ computed using Equation~(\ref{eq:scattering-force-op}) leads to the same expression given in Equation~(\ref{eq:force-res-singl}). Once the quantity $\langle \hat{\mathcal{F}}(t)\hat{\mathcal{F}}(t') \rangle$ has been written in terms of expectation values of the form $\langle \hat{a}^{\dagger}_{l_1}(E_1)\hat{a}_{l_2}(E_2)\hat{a}^{\dagger}_{l_3}(E_3)\hat{a}_{l_4}(E_4)\rangle$, using the Wick's theorem (see appendix \ref{app: B} for details) and performing the Fourier transform we get the final expression:
\begin{equation}
\label{eq:k0-int}
K_{0}=\frac{\hbar E_{F}}{\pi \delta^2}\int_{\frac{v_{m}}{v_F}}^{\infty}d \xi \frac{4 \alpha^{4} \xi^{-1}}{\xi^{2}+\alpha^{2}}\frac{1}{1+\cosh(\frac{1-\xi^2}{\tau})},
\end{equation}
where we have introduced the dimensionless temperature $\tau=k_{B}T/E_{F}$, an effective barrier width $\delta=2a$, the dimensionless barrier strength $\alpha=U \delta/(\hbar v_F)$ and the dimensionless particle velocity $\xi$ measured with respect to the Fermi velocity $v_F$. At low temperature the integral expression given in Equation~(\ref{eq:k0-int}) is well approximated by the following formula:
\begin{equation}
K_{0} \approx \frac{8 k_{B}T}{h v_{F}^{2}}\frac{U^{4}}{U^2+(\frac{\hbar v_F}{2a})^2},
\end{equation}
where $K_{0}$ is measured in units of $N^2/Hz$. Interestingly, both $\langle \hat{\mathcal{F}} \rangle$ and $K_{0}$ present a linear dependence on the reflection probability (computed at the Fermi energy $E_F$)
\begin{equation}
|r(E=E_F)|^2=\frac{U^2}{U^2+(\frac{\hbar v_F}{2a})^2}
\end{equation}
of the local potential $V(x)$. From a physical viewpoint, the force fluctuations described by $K_{0}$ are originated by thermal activated density fluctuations at the left and right side of the impurity site, determining a stochastic pressure imbalance on the barrier.
\subsection{Two-scatterer case}
\label{sec:force-two}
\begin{figure}
\includegraphics[clip,scale=0.435]{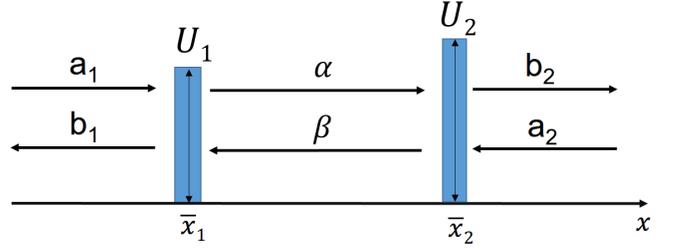}
\caption{(color online) Two-scatterer potential $V(x)=2a U_{1}\delta (x-\bar{x}_1)+2a U_{2}\delta (x-\bar{x}_2)$. Incoming ($a_i$) and outgoing ($b_i$) scattering operators are used to decompose the fermionic field $\hat{\Psi}(x)$ on the left ($x<\bar{x}_1$, $i=1$) and on the right ($x>\bar{x}_2$, $i=2$) side of the two-scatterer system. Middle scattering operators $\alpha$ and $\beta$ are used to describe the fluid confined between the two scatterers. The presence of a system edge in $x=\bar{x}_2$ can be studied considering the limit $U_2 \rightarrow \infty $ and restricting the analysis to $x<\bar{x}_2$. }
\label{fig:fig2}
\end{figure}

\begin{figure}
\includegraphics[clip,scale=0.85]{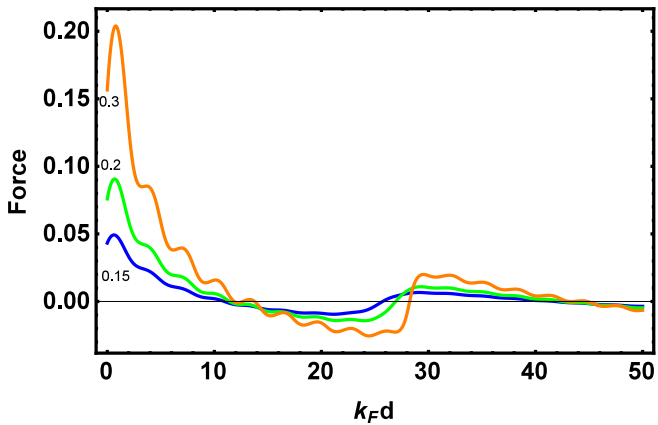}
\caption{(color online) Casimir force $\langle \hat{\mathcal{F}}_1\rangle$, in units of $E_F/(2 \pi a)$, acting in $\bar{x}_1$ as a function of the normalized scatterer distance $k_F d$, with $d=\bar{x}_2-\bar{x}_1$. Different curves are obtained considering the zero-temperature limit ($T=0$) and impurities with identical dimensionless scattering strength $\alpha_1=\alpha_2=\alpha$ ($\alpha=2aU/(\hbar v_F)$). The bottom curve is obtained fixing $\alpha=0.15$, middle and top curves are obtained fixing $\alpha=0.2$ and $\alpha=0.3$, respectively. Casimir force acting on the first barrier is attractive at short distance $k_F d \lesssim 12$, while a decaying attractive/repulsive behavior characterizes the long distance ($k_F d > 12$) impurity interaction. For the symmetric case ($\alpha_1=\alpha_2$), we have $\langle \hat{\mathcal{F}}_1\rangle=-\langle \hat{\mathcal{F}}_2\rangle$.}
\label{fig:fig3}
\end{figure}

\begin{figure}
\includegraphics[clip,scale=0.85]{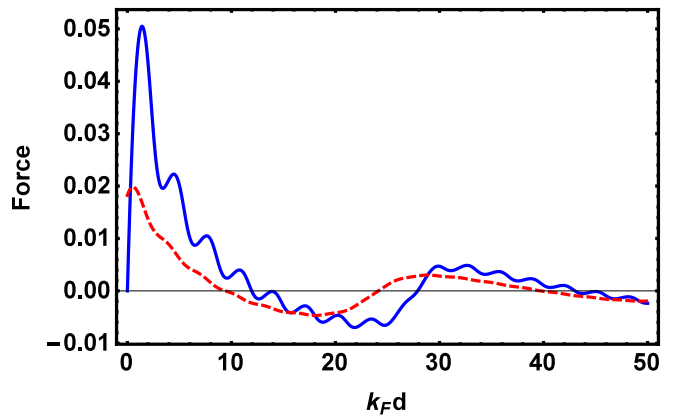}
\caption{(color online) Casimir force $\langle \hat{\mathcal{F}}_1\rangle$, in units of $E_F/(2 \pi a)$, acting in $\bar{x}_1$ as a function of the normalized scatterer distance $k_F d$, with $d=\bar{x}_2-\bar{x}_1$. The dashed curve (red in color) represents the Casimir interaction on the scatterer located in $\bar{x}_1$ in the presence of an identical scatterer located in $\bar{x}_2$, both having a dimensionless scattering strength $\alpha=0.1$. The full line (blue in color) represents the Casimir interaction on the scatterer located in $\bar{x}_1$ in the presence of an edge simulated by a second impurity of infinite scattering strength ($\alpha_2 \rightarrow \infty$). Zero-temperature limit is considered.}
\label{fig:fig4}
\end{figure}

In the following we study the two-scatterer problem under the assumption of ultralocal potentials explained before. The single particle potential produced by the scatterers located at $x=\bar{x}_1$ and $x=\bar{x}_2$, respectively, is thus given by $V(x)=2a U_{1}\delta (x-\bar{x}_1)+2a U_{2}\delta (x-\bar{x}_2)$. Different barrier strengths $U_{1,2}$ are allowed in the model (see Fig.~\ref{fig:fig2}). Our purpose is to study the force expectation value and its fluctuation acting on the scatterer located in $\bar{x}_1$ in the presence of a second distant scatterer located in $\bar{x}_2$.\\
To this end, we define the scattering fields $\hat{\Psi}_1(x,t)$, $\hat{\Psi}_m(x,t)$ and $\hat{\Psi}_2(x,t)$ describing fermions within the regions $x<\bar{x}_1$, $\bar{x}_1<x<\bar{x}_2$ and $x>\bar{x}_2$, respectively,
\begin{eqnarray}
\label{eq:scattering-fields-two}
&&\hat{\Psi}_{1}(x,t)=\int \frac{dE e^{-i\frac{Et}{\hbar}}}{\sqrt{h v(E)}}[\hat{a}_1(E) e^{i k_E x}+\hat{b}_1(E) e^{-i k_E x}]\nonumber \\
&&\hat{\Psi}_{m}(x,t)=\int \frac{dE e^{-i\frac{Et}{\hbar}}}{\sqrt{h v(E)}}[\hat{\alpha}(E) e^{i k_E x}+\hat{\beta}(E) e^{-i k_E x}]\nonumber\\
&&\hat{\Psi}_{2}(x,t)=\int \frac{dE e^{-i\frac{Et}{\hbar}}}{\sqrt{h v(E)}}[\hat{a}_2(E) e^{-i k_E x}+\hat{b}_2(E) e^{i k_E x}]\nonumber.
\end{eqnarray}
Here the scattering matrix $S_{ij}(E)$ of the two-impurity system is employed to write the relation $\hat{b}_{i}(E)=\sum_{j}S_{ij}(E) \hat{a}_{j}(E)$, while the dependence of the inner mode operators $\hat{\alpha}(E)$ and $\hat{\beta}(E)$ on $\hat{a}_i(E)$ can be expressed in the form:
\begin{eqnarray}
 \left(
 \begin{array}{c}
\hat{\alpha}(E) \\
  \hat{\beta}(E)
\end{array}
\right)=\mathcal{M}(E)\left(
 \begin{array}{c}
\hat{a}_1(E)\\
  \hat{a}_2(E)
\end{array}
\right),
\end{eqnarray}
where the explicit form of $\mathcal{M}(E)$ is given in Appendix~\ref{app: D}. The operators $\hat{\mathcal{F}}_{i}$ related with the force acting on the \textit{i}th scatterer can be written in terms of the scattering fields $\hat{\Psi}_{1,m,2}(x,t)$, complemented by Eq.~(\ref{eq:scattering-fields-two}), as specified below:
\begin{eqnarray}
\label{eq:scattering-force-bouble}
\hat{\mathcal{F}}_1=U_{1}\Big[\hat{\Psi}_{1}^{\dagger}(\bar{x}_1)\hat{\Psi}_{1}(\bar{x}_1)-\hat{\Psi}_{m}^{\dagger}(\bar{x}_1)\hat{\Psi}_{m}(\bar{x}_1)\Big]\nonumber \\
\hat{\mathcal{F}}_2=U_{2}\Big[\hat{\Psi}_{m}^{\dagger}(\bar{x}_2)\hat{\Psi}_{m}(\bar{x}_2)-\hat{\Psi}_{2}^{\dagger}(\bar{x}_2)\hat{\Psi}_{2}(\bar{x}_2)\Big],
\end{eqnarray}
while we have to characterize the force expectation value $\langle \hat{\mathcal{F}}_1 \rangle$ and its low-frequency fluctuation expressed via $K^{(1)}_0$.\\ As a preliminary task we present a qualitative description of the non-equilibrium force $\langle \hat{\mathcal{F}}_i \rangle$ acting on the \textit{i}th scatterer for a symmetric ($U_1=U_2=U$) system. Proceeding as described for the single scatterer case (Sec. \ref{sec:force-single}), we get the following result ($i \in \{1,2\}$)
\begin{eqnarray}
\label{eq:casimir1}
&&\langle \hat{\mathcal{F}}_i \rangle=\frac{U}{h}\int \frac{dE}{v(E)}|r(E)|^2(f_{1}(E)-f_{2}(E))+ \\
&&+(-)^{i} \frac{U}{h}\sum_{j}\int \frac{dE}{v(E)}[|\mathcal{M}_{1j}(E)|^2+|\mathcal{M}_{2j}(E)|^2-1]f_{j}(E)\nonumber,
\end{eqnarray}
where the first term represents the non-equilibrium force exerted by the electron wind on the \textit{i}th scatterer, while the second term is a non-equilibrium Casimir-like force contribution. Differently from the first term, the Casimir force provides a non-vanishing contribution also under equilibrium condition (i.e. $f_{1}(E)=f_{2}(E)$). Moreover the total force $\sum_{i}\langle \hat{\mathcal{F}}_i \rangle$ acting on the two-impurity system presents the same form of Equation~(\ref{eq:force-res-singl}), with $|r(E)|^2$ the reflection probability of the system containing two scatterers.\\
Under equilibrium condition, the two scatterers are only affected by the mechanical action provided by the Casimir force. In the following we characterize the Casimir force by studying $\langle \hat{\mathcal{F}}_{1}\rangle$ under equilibrium also considering asymmetric systems ($U_1 \neq U_2$). Under these assumptions  $\langle \hat{\mathcal{F}}_{1}\rangle$ can be written as:
\begin{eqnarray}
\langle \hat{\mathcal{F}}_{1}\rangle=\frac{U_1}{h}\sum_{j}\int \frac{dE}{v(E)}[1-|\mathcal{M}_{1j}(E)|^2-|\mathcal{M}_{2j}(E)|^2]f(E) \nonumber ,
\end{eqnarray}
where the information on the impurity distance\cite{note2} $d=\bar{x}_2-\bar{x}_1$ and on the barrier strengths, namely $U_1$ and $U_2$, is encoded within the M-matrix elements $\mathcal{M}_{ij}(E)$.\\
Hereafter, we provide a numerical study of the Casimir force adopting dimensionless barrier strengths $\alpha_i= 2a U_i/(\hbar v_F)$, dimensionless impurity distance $k_F d$ and measuring the Casimir force in units of $E_F/(2\pi a)$. For a typical nanostructured system, we assume $m=0.1 \times m_0$ ($m_0$, being the bare electron mass), $v_F = 10^6 m/s$, $a=0.1$ nm, and thus the force unit $E_F/(2\pi a)$ is about $72.5$ pN, which represents an intense interaction on the nanoscale. In Fig.~\ref{fig:fig3} we study the Casimir force $\langle \hat{\mathcal{F}}_1\rangle$ acting on the scatterer located in $\bar{x}_1$ as a function of the normalized scatterer distance $k_F d$ ($d=\bar{x}_2-\bar{x}_1$). The numerical analysis shows that the Casimir force is weakly affected by temperature provided that $k_B T/E_F \approx 10^{-3} \div 10^{-2}$ (low-temperature limit) and thus we consider the zero-temperature limit. In Fig.~\ref{fig:fig3} a system made of two identical impurities with dimensionless scattering strength $\alpha_1=\alpha_2=\alpha$ is considered. The bottom curve is obtained fixing $\alpha=0.15$, middle and top curves are obtained fixing $\alpha=0.2$ and $\alpha=0.3$, respectively. The Casimir force acting on the first barrier presents an attractive character at short distance $k_F d \lesssim 12$, while a decaying attractive/repulsive behavior characterizes the long distance ($k_F d > 12$) impurity interaction. Furthermore, for the symmetric case ($\alpha_1=\alpha_2$), we have $\langle \hat{\mathcal{F}}_1\rangle=-\langle \hat{\mathcal{F}}_2\rangle$. The sign changing of the Casimir force as function of the scatterer distance $d$ is a coherence effect which is reminiscent of the Friedel density oscillations of the quantum fluid. The basic properties of these force oscillations can be captured by considering the limit of very small scattering strength $\alpha$. Under this condition, introducing a low-velocity cutoff $v_m$ and taking the zero-temperature limit we obtain:
\begin{equation}
\langle \hat{\mathcal{F}}_{1}\rangle \propto \int_{\frac{v_m}{v_F}}^{1} d\xi \frac{4 \alpha^3 \cos(2 k_F d \ \xi)}{\xi^2},
\end{equation}
being this expression valid within the symmetric case $\alpha_1=\alpha_2=\alpha$ and provided that $\alpha \lesssim 1.5 \times 10^{-2}$. Thus quantum coherence seems to be a crucial ingredient in observing a non-vanishing mechanical action mediated by a quantum fluid. Indeed, the Casimir energy associated to the impurity interaction mediated by a coherent one-dimensional fluid takes the asymptotic form\cite{recatiPRA05} $E_{c}(x) \propto \mathcal{W}(x)/x$, with $\mathcal{W}(x)$ an oscillating function of the impurity distance $x$ related to the Friedel density oscillations. Using $E_{c}(x)$, the Casimir force takes the form $F_c(x)=-\partial_x E_{c}(x) \propto x^{-2} \mathcal{W}(x)-x^{-1}\partial_x \mathcal{W}(x)$ which defines a long-range interaction which decays as the inverse of the impurities distance ($F_{c} \sim x^{-1}$). We verified the long-distance behavior of the $\langle \hat{\mathcal{F}}_1\rangle$ \textit{vs} $k_F d$  curves reported in Fig.~\ref{fig:fig3} and a $d^{-1}$ scaling is found. This long-range behavior suggests that the Casimir force has an enhanced relevance in one-dimensional systems where it could be easily detected in comparison with the three-dimensional case.\\
Another important point is the existence of a non-vanishing Casimir force acting on an impurity in close proximity of an edge of the system. The presence of an edge in $x=\bar{x}_2$ can be emulated by a second scatterer of infinite scattering strength ($\alpha_2 \rightarrow \infty$) and considering the resulting system for $x<\bar{x}_2$. This scenario is studied in Fig.~\ref{fig:fig4} where the Casimir force exerted by an edge (full line) on the impurity located in $\bar{x}_1$ is compared to the Casimir force acting on the same impurity and originated by an identical scatterer (dashed line) having scattering strength $\alpha=0.1$. As clearly shown in Fig.~\ref{fig:fig4}, the system edge provides a stronger Casimir force compared to the one generated by an identical scatterer placed at the same distance $k_F d$. Thus, the edge mediated Casimir force provides a mechanism of eliminating the phase defects (impurities) confining them towards the system boundary. This observation suggests that the short-distance edge-mediated Casimir attraction can hinder the diffusion towards the system's bulk of atomic dopants (impurities) located in close vicinity of the system boundaries.\\
So far we noticed that the Casimir force $\langle \hat{\mathcal{F}}_1\rangle$ presents only a very weak temperature dependence coming from the Fermi function in Eq.~(\ref{eq:casimir1}); we also noticed that the Casimir forces are originated by the coherence of the quantum fluid as signaled by Friedel-like oscillations. In this respect, low-temperature are crucial in obtaining coherence and thus non-vanishing Casimir forces. To support these arguments, we introduce a random (gaussian) dephasing $\phi$ within the scattering region mimicking decoherence phenomena induced by inelastic scattering events (see Appendix \ref{app: C} for details). Using the scattering field theory we derive the phase-dependent Casimir force $\langle \hat{\mathcal{F}}_1\rangle_{\phi}$, which has to be averaged over random realizations of inelastic events using the gaussian probability density function $P(\phi,\sigma^2)$, i.e. $\overline{\langle \hat{\mathcal{F}}_1\rangle_{\phi}}=\int d \phi \langle \hat{\mathcal{F}}_1\rangle_{\phi}P(\phi,\sigma^2)$. The Casimir force under incoherent regime $\overline{\langle \hat{\mathcal{F}}_1\rangle_{\phi}}$ depends on the variance $\sigma^2$ of the phase fluctuation within the scattering region, $\sigma^2$ being a decoherence measure. The dephasing variance, indeed, can be related to the coherence length $L_{\Phi}$ of the system through the relation $\sigma^2=L/L_{\Phi}$, where $L$ is a mesoscopic distance comparable with the scattering region length $d$. The coherence length $L_{\Phi}$ decreases as the temperature is increased and thus the dephasing variance $\sigma^2$ takes high values and induces decoherence when $L>L_{\Phi}$. On the other hand, the coherent regime is recovered at low temperature where the coherence length $L_{\Phi}$ becomes larger than the scattering region length $d \sim L$, while $\sigma \rightarrow 0$. According to these arguments, the detrimental effects of temperature affect the Casimir force through the parametric dependence of the coherence length $L_{\Phi}(T)$ on temperature. Thus Casimir mediated forces disappear at sufficiently high temperature.\\
\begin{figure}
\includegraphics[clip,scale=0.85]{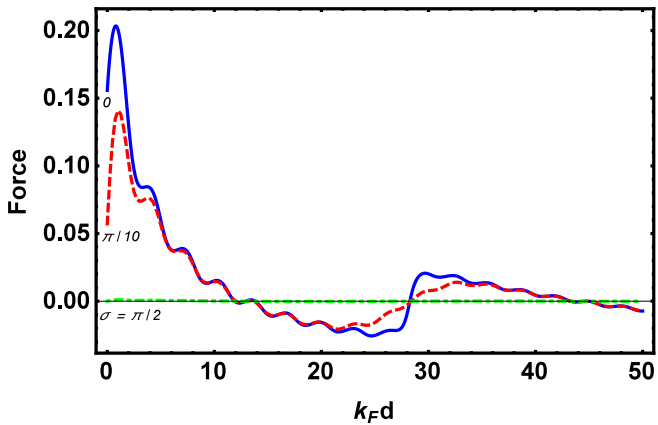}
\caption{(color online) Casimir force $\langle \hat{\mathcal{F}}_1\rangle$, in units of $E_F/(2 \pi a)$, acting in $\bar{x}_1$ as a function of the normalized scatterer distance $k_F d$, with $d=\bar{x}_2-\bar{x}_1$. The full line curve (blue in color) represents the coherent ($\sigma=0$) Casimir interaction on the scatterer located in $\bar{x}_1$ in the presence of an identical scatterer located in $\bar{x}_2$, both having a dimensionless scattering strength $\alpha=0.3$. Decoherence is introduced in the system by considering a gaussian random dephasing with standard deviation $\sigma=\pi/10$ (dashed curve, red in color) and $\sigma=\pi/2$ (dashed dotted curve, green in color). Zero-temperature limit is considered. The Casimir interaction is strongly suppressed as the decoherence increases.}
\label{fig:fig5}
\end{figure}
In Fig.~\ref{fig:fig5} we study the effect of decoherence on the Casimir force $\langle \hat{\mathcal{F}}_1\rangle$ acting on the impurity located in $x=\bar{x}_1$ in the presence of a second identical impurity having scattering strength $\alpha=0.3$ and located in $x=\bar{x}_2$. The full line curve (blue in color) represents the coherent ($\sigma=0$) Casimir interaction, while decoherence is progressively introduced considering random dephasing with standard deviation $\sigma=\pi/10$ (dashed curve, red in color) and $\sigma=\pi/2$ (dashed dotted curve, green in color). We notice that the Casimir force disappears as $\sigma$ is increased. In particular, the choice $\sigma=\pi/10$ determines a coherence length $L_{\Phi}$ twenty-five times bigger than the coherence length obtained for $\sigma=\pi/2$. As a consequence, in the latter case, the dimensionless Casimir force is always weaker than $10^{-3}$. Thus Casimir interaction can be neglected at high temperature, i.e. when the coherence length is small compared to the inter-impurity distance $d$. Decreasing the temperature, the coherence length starts to become greater than $d$ and a non-local interaction mediated by the coherent fluid affects the impurity dynamics. Under this condition, the vibrations of close impurities are correlated by the Casimir interaction. However, in a condensed matter system defects are pinned at specific lattice sites as the effect of elastic forces. This implies that the equilibration of Casimir force with elastic deformations of the lattice determines a new equilibrium position of the impurity accompanied by a local mechanical stress. The quantum origin of the Casimir interaction determines correlated force fluctuations on the impurities. These fluctuations affect the stochastic impurity vibrations around the equilibrium positions and can play a role in determining temporal variations of the system electrical resistance. When a conductive (ohmic) system is subject to a current bias, resistance fluctuations determine stochastic voltage variations which can be studied using the voltage noise spectroscopy. In these studies, the voltage spectral density presents a $1/f$-spectrum and a quadratic dependence on the bias current. The latter phenomenology, which is quite common in metals, is explained as the cumulative effect of uncorrelated stochastic fluctuators (impurities) having lorentzian spectrum. Recently, it has been reported that a large class of materials undergoing a weak localization (WL) transition exhibits a sudden variation of the voltage spectral density at the  WL transition temperature \cite{wl-noise}. This phenomenon has been attributed to the partial restoration of the particle coherence in WL regime. These findings suggest that fluid-mediated force fluctuations can affect the stochastic properties of fluctuators introducing unexpected correlation effects among vibrations of close impurities. Motivated by these arguments, we studied the low-frequency force fluctuation $K_{0}^{(1)}$ acting on the first scatterer ($x=\bar{x}_1$) in the presence of a second one located in $x=\bar{x}_2$. Since we are interested in the coherence properties of the system, low-temperature and coherent regime ($\sigma=0$) are assumed in the computation. The derivation of $K_{0}^{(1)}$ closely follows the procedure adopted in the single impurity case and thus in the following we only describe the numerical results.\\
In particular, in Fig.~\ref{fig:fig6} we study the fluctuation $K_{0}^{(1)}$ (in units of $\hbar E_F/(\pi \delta^{2})$) acting on the first scatterer ($x=\bar{x}_1$) in the presence of a second one located in $x=\bar{x}_2=\bar{x}_{1}+d$. The lower curve (red in color) is obtained by fixing the model parameters as: $\alpha_1=\alpha_2=0.7$, $k_B T =10^{-2}E_{F}$. It shows peaks associated with the transmission resonances of the two-impurity system as the impurity distance $d$ is varied. The fluctuation peaks determine an increased noise level compared to the value expected for an isolated impurity having the same scattering strength (i.e. for $\alpha=0.7$ and $k_B T =10^{-2}E_{F}$).
\begin{figure}
\includegraphics[clip,scale=0.85]{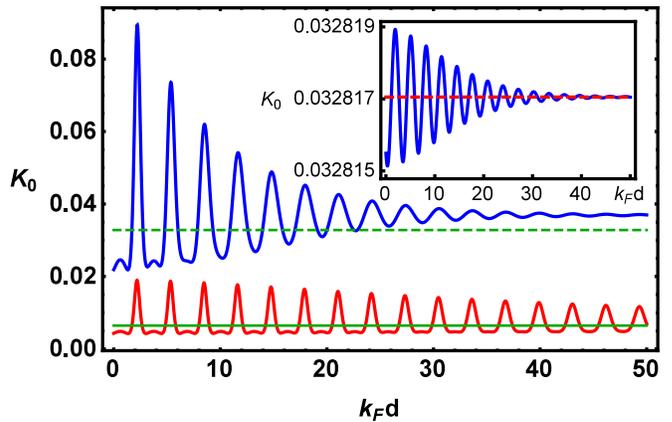}
\caption{(color online) Low-frequency force fluctuation $K_{0}^{(1)}$ (in units of $\hbar E_F/(\pi \delta^{2})$) acting on the first scatterer ($x=\bar{x}_1$) in the presence of a second one located in $x=\bar{x}_2=\bar{x}_{1}+d$. Lower curve (red in color) is obtained by fixing the model parameters as: $\alpha_1=\alpha_2=0.7$, $k_B T =10^{-2}E_{F}$. Upper curve (blue in color) is obtained by fixing the model parameters as: $\alpha_1=\alpha_2=0.7$, $k_B T =5 \times 10^{-2}E_{F}$. The curve in the inset is obtained setting the parameters as: $\alpha_{1}=0.7$, $\alpha_{2}=5 \times 10^{-5}$, $k_{B}T=5 \times 10^{-2}E_F$. Parallel lines (green lines in the main figure and red line in the inset) indicate the noise level expected for an isolated impurity (Eq.~(20)) with the same scattering strength $\alpha$ of the impurity in $x=\bar{x}_1$ at the same temperature. Peaks in the fluctuations are related to the transmission resonances of the two-impurity system.}
\label{fig:fig6}
\end{figure}
This behavior persists for large separation between the impurities even though the fluctuation peak amplitudes decrease with the distance $d$. Interestingly, the Casimir force fluctuations decrease slower than the Casimir force expectation value as the scatterer distance is increased. Thus, also in the absence of a net Casimir force, Casimir force fluctuations can play a relevant role in coupling the dynamics of distant scatterers. The upper curve (blue in color) is obtained by increasing the system temperature ($k_B T= 5 \times 10^{-2} E_F$), while setting the scattering strength of the impurities as done for the lower curve. The higher system temperature amplifies the mean fluctuation level (and the fluctuation peak amplitudes) and provides an averaging mechanism which produces a long-distance constant value of the fluctuation. The long-distance fluctuation value is higher than the noise level of an isolated impurity (dashed green line) since the system reflection probability of the single- and two-impurity cases are different. The single-impurity value (dashed red line of the inset) of the fluctuation is recovered at long-distance when the scattering strength of the second impurity is strongly reduced ($\alpha_{2}=5 \times 10^{-5}$), as shown in the inset.\\
Finally, we conclude that quantum force fluctuations acting on a system impurity can be strongly amplified (compared to the isolated impurity case) in the presence of a second impurity. This implies that the mechanical actions affecting impurities immersed in a coherent quantum fluid are correlated via a Casimir-like mechanism which can play a relevant role in low-dimensional systems.

\section{Conclusions}
\label{sec:concl}
We have formulated a scattering field theory accounting for the mechanical actions experienced by impurities located in a coherent fermionic fluid. The theory allows the study of equilibrium and out-of-equilibrium situations taking into account the impurity-fluid interaction at a non-perturbative level. Using the theory, we studied the single-impurity and the two-impurity cases. As for the single-scatterer case, we find that a net force affects the scatterer dynamics only in the presence of a non-vanishing particle current flowing through the system (non-equilibrium), while force fluctuations are not negligible both in equilibrium and in non-equilibrium conditions. Concerning the two-scatterer case, an attractive fluid-mediated Casimir force is experienced by the scatterers at small spatial separation, while a decaying attractive/repulsive behavior as a function of the scatterer distance is found. We also find a Casimir-like force experienced by an impurity in close vicinity of the boundary of the system. For the one dimensional system studied in this work, the Casimir force decays as the inverse of the impurity distance and thus represents a long-range interaction which can be easily measured, for instance, in cold atoms systems. When detrimental effects of decoherence are included in the model, the Casimir interaction is progressively reduced as the system coherence length $L_{\Phi}$ is decreased. We also studied the Casimir force fluctuations, measured by $K_{0}^{(1)}$, acting on a given scatterer in close vicinity of an other one. We show that the $K_{0}^{(1)}$ \textit{vs} $k_F d$ curves present an oscillating behavior reaching a long-distance limit which is comparable with the value of the single-scatterer case. Depending on the impurity separation, at short distance, force fluctuations can became either stronger or weaker than the force variance experienced by an isolated impurity. Our findings demonstrate that force fluctuations acting on close impurities immersed in a coherent fluid are correlated by Casimir-like interactions which are reminiscent of Friedel density oscillations and resonant scattering through the double-impurity system.

\section*{Acknowledgements}
Antonio Stabile is acknowledged for inspiring discussions about Casimir effect. Discussions with Roberta Citro, Antonio Di Bartolomeo, Filippo Giubileo, Carlo Barone and Sergio Pagano are also acknowledged.

\appendix

\section{S-matrix and T-matrix of the scattering potential $V(x)=\lambda \delta (x-x_0)$}
\label{app: A}
Using the Schr\"{o}dinger-like equation for the field $\hat{\Psi}(x)$ with the single-particle potential $V(x)=\lambda \delta (x-x_0)$ the following boundary conditions on $\hat{\Psi}(x)$ can be found:
\begin{eqnarray}
\label{eq:BcsApp}
&&\partial_x \hat{\Psi}(x)|_{x=x_0^{+}}-\partial_x \hat{\Psi}(x)|_{x=x_0^{-}}=\frac{2 m \lambda}{\hbar^2}\hat{\Psi}(x_0) \nonumber \\
&&\hat{\Psi}(x_0^{+})=\hat{\Psi}(x_0^{-}).
\end{eqnarray}
The field $\hat{\Psi}(x)$ on the left ($x<x_0$) and on the right ($x>x_0$) of the scatterer can be decomposed in incoming ($\hat{a}_i$) and outgoing ($\hat{b}_i$) scattering operators according to the following expansion:
\begin{eqnarray}
\label{eq:wfsApp}
\hat{\Psi}_{E}(x<x_0)=\hat{a}_1(E) e^{i k_{E}x}+\hat{b}_1(E) e^{-i k_{E}x} \nonumber \\
\hat{\Psi}_{E}(x>x_0)=\hat{a}_2(E) e^{-i k_{E}x}+\hat{b}_2(E) e^{i k_{E}x}.
\end{eqnarray}
Substituting Eq.~(\ref{eq:wfsApp}) into Eq.~(\ref{eq:BcsApp}) one can find a relation between the scattering operators $\hat{a}_i(E)$ and $\hat{b}_{i}(E)$. This relation can be expressed in terms of the S-matrix according to the expression $\hat{b}_{i}(E)=\sum_{j}S_{ij}(E)\hat{a}_{j}(E)$, while the T-matrix relation takes the following form:
\begin{eqnarray}
 \left(
 \begin{array}{c}
\hat{b}_2(E) \\
  \hat{a}_2(E)
\end{array}
\right)=T(E)\left(
 \begin{array}{c}
\hat{a}_1(E)\\
  \hat{b}_1(E)
\end{array}
\right).
\end{eqnarray}
For the potential under discussion we obtain the following expression for S- and T-matrix:
\begin{eqnarray}
S(E)=\left(\begin{array}{cc}
          \frac{\gamma_{E}e^{2 i k_{E}x_0}}{2i-\gamma_{E}} &  \frac{2}{2i+\gamma_{E}} \\
          \frac{2}{2i+\gamma_{E}} & \frac{\gamma_{E}e^{-2 i k_{E}x_0} }{2i-\gamma_{E}}
        \end{array}
\right)
\end{eqnarray}
and
\begin{eqnarray}
T(E)=\frac{1}{2}\left(\begin{array}{cc}
          2-i \gamma_{E} & -i\gamma_{E} e^{-2 i k_{E} x_0} \\
          i\gamma_{E} e^{2 i k_{E} x_0} & 2+i \gamma_{E}
        \end{array}
\right)
\end{eqnarray}
with $\gamma_{E}=2m\lambda/(\hbar^2 k_{E})$ and $E=(\hbar k_E)^2/(2m)$. In general, it is possible to demonstrate that $det(T)=1$, while under time-reversal symmetry the S-matrix elements are related to those of the T-matrix as follows: $S_{12}=S_{21}=1/T_{22}$, $S_{11}=-T_{21}/T_{22}$, $S_{22}=T_{12}/T_{22}$.

\section{Thermal averages of scattering fields}
\label{app: B}
The quantum statistical average of two scattering operators is given by $\langle \hat{a}^{\dagger}_{i}(E)\hat{a}_{j}(E')\rangle=\delta_{ij}\delta(E-E')f_{i}(E)$. Quantum averages involving four scattering operators can be evaluated using Wick's theorem obtaining the following result:
\begin{eqnarray}
&&\langle \hat{a}^{\dagger}_{l_1}(E_1)\hat{a}_{l_2}(E_2)\hat{a}^{\dagger}_{l_3}(E_3)\hat{a}_{l_4}(E_4)\rangle= \\
&& \delta_{l_1,l_2}\delta_{l_3,l_4} \delta(E_1-E_2)\delta(E_3-E_4)f_{l_1}(E_1)f_{l_3}(E_3)+\nonumber\\
&&\delta_{l_1,l_4}\delta_{l_2,l_3} \delta(E_1-E_4)\delta(E_2-E_3)f_{l_1}(E_1)[1-f_{l_3}(E_3)]\nonumber.
\end{eqnarray}
In deriving the above relation we used  the anticommutation relation $\{\hat{a}_{l_1}(E_1),\hat{a}^{\dagger}_{l_2}(E_2)\}=\delta_{l_1,l_2}\delta(E_1-E_2)$.

\section{$\mathcal{M}(E)$ Matrix}
\label{app: D}
The matrix $\mathcal{M}(E)$ defining the relation between the inner mode operators $\hat{\alpha}(E)$ and $\hat{\beta}(E)$ on $\hat{a}_i(E)$ can be written in terms of the transfer matrix $T_1(E)$ of the impurity located in $x=\bar{x}_1$ and of the scattering matrix elements $S_{ij}(E)$ of the whole system according to the relation:
\begin{eqnarray}
\label{eq: m-matrix}
\mathcal{M}(E)=T_1(E) \cdot \left(\begin{array}{cc}
          1 & 0 \\
          S_{11}(E) & S_{12}(E)
        \end{array}
\right).
\end{eqnarray}
The derivation of the above result exploits the following relations linking the scattering fields:
\begin{eqnarray}
 \left(
 \begin{array}{c}
\hat{\alpha}(E) \\
  \hat{\beta}(E)
\end{array}
\right)=T_1(E)\left(
 \begin{array}{c}
\hat{a}_1(E)\\
  \hat{b}_1(E)
\end{array}
\right),
\end{eqnarray}

\begin{eqnarray}
 \left(
 \begin{array}{c}
\hat{b}_2(E) \\
  \hat{a}_2(E)
\end{array}
\right)=T_2(E)\left(
 \begin{array}{c}
\hat{\alpha}(E)\\
  \hat{\beta}(E)
\end{array}
\right),
\end{eqnarray}

\begin{eqnarray}
 \left(
 \begin{array}{c}
\hat{b}_2(E) \\
  \hat{a}_2(E)
\end{array}
\right)=T(E)\left(
 \begin{array}{c}
\hat{a}_1(E)\\
  \hat{b}_1(E)
\end{array}
\right),
\end{eqnarray}
where $T(E)=T_2(E) \cdot T_1(E)$ represents the transfer matrix of the two-impurity system written in terms of the transfer matrices of the first and second impurity. The matrix $\mathcal{M}(E)$ can be also derived in terms of $T_2(E)$ according to the expression:
\begin{eqnarray}
\mathcal{M}(E)=T^{-1}_2(E) \cdot \left(\begin{array}{cc}
          S_{21}(E) & S_{22}(E) \\
          0 & 1
        \end{array}
\right),
\end{eqnarray}
the latter formula providing the same result of Equation~(\ref{eq: m-matrix}).

\section{T-matrix model of random dephasing}
\label{app: C}
Dephasing effects (decoherence) can be studied by introducing random phases within the scattering region \cite{decoherence_refs}. In particular we consider the scattering matrix $S_{\phi}$ describing a pure (reflectionless) dephasing process:
\begin{eqnarray}
S_{\phi}=e^{i \phi}\left( \begin{array}{cc}
                            0 & 1 \\
                            1 & 0
                          \end{array}
\right),
\end{eqnarray}
where $\phi$ represents a random gaussian variable (with probability density distribution $P(\phi, \sigma^2)=\exp(-\phi^2/(2\sigma^2))/\sqrt{2 \pi \sigma^2}$) having zero expectation value, $\overline{\phi}=0$, and finite variance $\sigma^2=\overline{\phi^2}$, the latter quantity being related with the decoherence. The scattering matrix $S_{\phi}$ determines the relations $\hat{b}_1=e^{i\phi} \hat{a}_2$ and $\hat{b}_2=e^{i\phi} \hat{a}_1$ such that the T-matrix of a pure dephasing process can be written as:
\begin{eqnarray}
\tau_{\phi}=\left( \begin{array}{cc}
                            e^{i \phi} & 0 \\
                            0 & e^{-i \phi}
                          \end{array}
\right).
\end{eqnarray}
Let us study a sequence of $N$ scatterers located at $x_1< x_2< ...< x_N$ and characterized by a collection of T-matrices whose generic element $T(x_i)$ describes the scattering properties of the \textit{i}th scatterer. Using the T-matrix composition rule, the T-matrix of the whole system is simply given by the matrix product $T=T(x_N)\ldots T(x_1)$. In the presence of decoherence effects the T-matrix composition rule can be corrected inserting the dephasing matrix, i.e. $\widetilde{T}=T(x_N)\ldots \tau_{\phi} \cdot T(x_1)$. Using $\widetilde{T}$ within the scattering field theory, we can compute any observable $X(\phi)$ at fixed dephasing $\phi$, while the decoherence effect on the observable is obtained performing the phase average $\overline{X(\phi)}=\int d\phi X(\phi) P(\phi,\sigma^2)$ using the probability density distribution of the phase fluctuations $P(\phi,\sigma^2)$. Assuming that the phase dependence of the physical observable can be expanded according to the trigonometric series $X(\phi)=\sum_{n=0}^{\infty}[c_n \cos(n \phi)+s_n \sin(n \phi)]$, we obtain the phase-averaged observable
\begin{equation}
\overline{X(\phi)}=c_0+\sum_{n=1}^{\infty}c_n \exp \Big(-\frac{n^2 \sigma^2}{2}\Big),
\end{equation}
showing the high harmonics smearing and a tendency to converge towards the average value $c_0$ as the degree of decoherence increases ($\sigma \rightarrow \infty$). In deriving the above result we used $\overline{\sin(n \phi)}=0, \forall n$ due to the even integration domain $]-\infty,\infty[$ of the $\phi$ variable.


\begin{thebibliography}{99}
\bibitem{casimir}H. B. G. Casimir, Proc. K. Ned. Akad. Wet. \textbf{51}, 793-795 (1948).
\bibitem{mostepanenko-book} V. Mostepanenko and N. Trunov, \textit{The Casimir Effect and Its Applications} (Clarendon, Oxford, 1997).
\bibitem{casimir-exp}S. K. Lamoreaux, Phys. Rev. Lett. \textbf{78}, 5 (1997); S. K. Lamoreaux, Phys. Rev. Lett. \textbf{81}, 5475 (1998); U. Mohideen and A. Roy, Phys. Rev. Lett. \textbf{81}, 4549 (1998); G. Bressi, G. Carugno, R. Onofrio, and G. Ruoso, Phys.
Rev. Lett. \textbf{88}, 041804 (2002); R. S. Decca, D. L\'{o}pez, H. B. Chan, E. Fischbach, D. E. Krause, and C. R. Jamell, Phys. Rev. Lett. \textbf{94}, 240401 (2005).

\bibitem{casimir-geometry} T. H. Boyer, Phys. Rev. \textbf{174}, 1764 (1968); T. Emig, A. Hanke, and M. Kardar, Phys. Rev. Lett. \textbf{87},
260402 (2001); O. Schroeder, A. Scardicchio, and R. L. Jaffe, Phys. Rev. A \textbf{72}, 012105 (2005); T. Emig, A. Hanke, R. Golestanian, and M. Kardar,
Phys. Rev. Lett. \textbf{95}, 250402 (2005).

\bibitem{capasso2009} J. N. Munday, Federico Capasso, and V. Adrian Parsegian, Nature \textbf{457}, 170 (2009).

\bibitem{zwerger2007} J. N. Fuchs, A. Recati, and W. Zwerger, Phys. Rev. A \textbf{75}, 043615 (2007);
P. W\"{a}chter, V. Meden, and K. Sch\"{o}nhammer, Phys. Rev. B \textbf{76}, 045123 (2007).
\bibitem{recatiPRA05} A. Recati, J. N. Fuchs, C. S. Pe\c{c}a, and W. Zwerger, Phys. Rev. A \textbf{72}, 023616 (2005).
\bibitem{zhabinskaya2008} D. Zhabinskaya, J. M. Kinder, and E. J. Mele, Phys. Rev. A \textbf{78}, 060103(R) (2008).


\bibitem{recati2005} A. Recati, P. O. Fedichev, W. Zwerger, J. von Delft, and P. Zoller, Phys. Rev. Lett. \textbf{94}, 040404 (2005).

\bibitem{ferone2010} Pawel Utko, Raffaello Ferone, Ilya V. Krive,	Robert I. Shekhter,	Mats Jonson, Marc Monthioux, Laure No\'{e} and Jesper Nyg{\aa}rd, Nat. Commun. \textbf{1}, 37 (2010).

\bibitem{nanotube-ads} Boris Dzyubenko,	Hao-Chun Lee, Oscar E. Vilches David H. Cobden, Nature Physics \textbf{11}, 398 (2015); F. Romeo, R. Citro, A. Di Bartolomeo, Phys. Rev. B \textbf{84}, 153408 (2011).

\bibitem{casimir-fluct2010}Anne-Florence Bitbol, Paul G. Dommersnes, and Jean-Baptiste Fournier, Phys. Rev. E \textbf{81}, 050903(R) (2010).

\bibitem{yurke90} B. Yurke and G. P. Kochanski, Phys. Rev. B \textbf{41}, 8184 (1990).
\bibitem{presilla92} C. Presilla, R. Onofrio and M. F. Bocko, Phys. Rev. B \textbf{45}, 3735 (1992).
\bibitem{buttiker92} M. B\"{u}ttiker, Phys. Rev. B \textbf{46}, 12485 (1992).
\bibitem{note1} In the spinful case a prefactor two accounting for the spin degeneracy has to be included.
\bibitem{note2} For consistency reasons, the force $\langle \hat{\mathcal{F}}_i\rangle$ has to be computed assuming impurity distance $d>2a$, being $a$ a small quantity of the theory. This constraint ensures that the scattering field $\hat{\Psi}_{m}(x,t)$ is a well-defined quantity acting on a finite length region.
\bibitem{wl-noise}C. Barone, F. Romeo, A. Galdi, P. Orgiani, L. Maritato, A. Guarino, A. Nigro, S. Pagano, Phys. Rev. B \textbf{87}, 245113 (2013); C. Barone, F. Romeo, S. Pagano, C. Attanasio, G. Carapella, C. Cirillo, A. Galdi, G. Grimaldi, A. Guarino, A. Leo, A. Nigro, P. Sabatino, Sci. Rep. \textbf{5}, 10705 (2015).
\bibitem{decoherence_refs} Marco Pala and Giuseppe Iannaccone, Phys. Rev. B \textbf{69}, 235304 (2004); Marco G. Pala and Giuseppe Iannaccone, Phys. Rev. Lett. \textbf{93}, 256803 (2004); Huaixiu Zheng, Zhengfei Wang, Qinwei Shi, Xiaoping Wang and Jie Chen, Phys. Rev. B \textbf{74}, 155323 (2006).
\end{thebibliography}
\end{document}